%% file: BESIIILHSforFPCP2018Proc.tex
\documentclass[graybox]{svmult}

\usepackage{mathptmx}       
\usepackage{helvet}         
\usepackage{courier}        
\usepackage{type1cm}        
\usepackage{makeidx}         
\usepackage{graphicx}        
\usepackage{multicol}        
\usepackage{bm}   
\usepackage[bottom]{footmisc}

\makeindex             

\begin{document}

\title*{Light Hadron Spectroscopy and Decay at BESIII}
\author{Vindhyawasini Prasad}
\institute{Vindhyawasini Prasad \at Department of Modern Physics, University of Science $\&$ Technology of China; State Key Laboratory of Particle Detection and Electronics, Hefei, 230026, Beijing 100049, China. \\
\email{vindy@ustc.edu.cn}
}
%
%
\maketitle

\abstract{Light hadron spectroscopy plays an important role in understanding the decay dynamics of unconventional hadronic states, such as strangeonium and glueballs. BESIII provides an ideal avenue to search for these exotic states thanks to a huge amount of data recorded at various energy points in the tau-charm mass region including  $J/\psi$ resonance. This report summarizes  recent results of the BESIII experiment related to the glueballs and strangeonium-like states. }

\section{Introduction}
\label{sec:1}
Quantum chromodynamics (QCD) describes hadrons as the bound states of quarks held together via the color force mediated by gluons. The non-Abelian nature of the QCD allows the self-interaction of gluons that can form the hadronic matter. These quarkless states are called \rm{\lq glueballs\rq}. The lattice QCD predicts that the glueballs having mass within the range of $1-2$ ($2-3$) GeV/$c^{2}$ are scalar (tensor or pseudoscalar)~\cite{latticeQCD}. The branching fractions for glueballs in the radiative decays of $J/\psi$ are expected to be within the range of $10^{-2}-10^{-3}$ depending upon the exact nature of  glueballs~\cite{reditJpsglueb}. The large data samples collected at the center-of-mass (CM) energies  between $2.0-4.6$ GeV, more than 130 energy points, including the $J/\psi$ resonance, by  BESIII provide an ideal avenue to explore the possibilities of these glueballs. The QCD also allows other possible forms of multiquark and hybrid states, such as a tetraquark-like $Z_c$ that was observed in the $\pi J/\psi$ mass spectrum in $\rm{Y}(4260) \to \pi \pi J/\psi$ decays by the BESIII~\cite{zc} and intermediately confirmed by Belle experiment~\cite{zcbelle}. Similar to $Z_c$~\cite{zc,zcbelle}, a strangeonium-like state is also expected  in the $\pi \phi$ spectrum via $\phi(2170) \to \pi\pi \phi$ decays, where $\phi(2170)$ [also denoted as Y(2175)] is analogous to Y(4260)~\cite{y4260}. The $\phi(2170)$ was observed by the BaBar collaboration via the initial-state-radiation process $e^+e^- \to \gamma \phi  f_0(980)$~\cite{babary2175}, and  later confirmed by Belle~\cite{belley2175}, BESII~\cite{bes2y2175} and BESIII collaborations~\cite{bes3y2175}. Following sections summarize  recent results of BESIII related to the gluonic and strangeonium-like states.

\section{Search for gluonic states}
\subsection{Anomalous enhancement at the $\bm{p\overline{\bm p}}$ mass-threshold}
\label{sec:2}
\begin{figure}[!htp]
\sidecaption
\includegraphics[scale=.62]{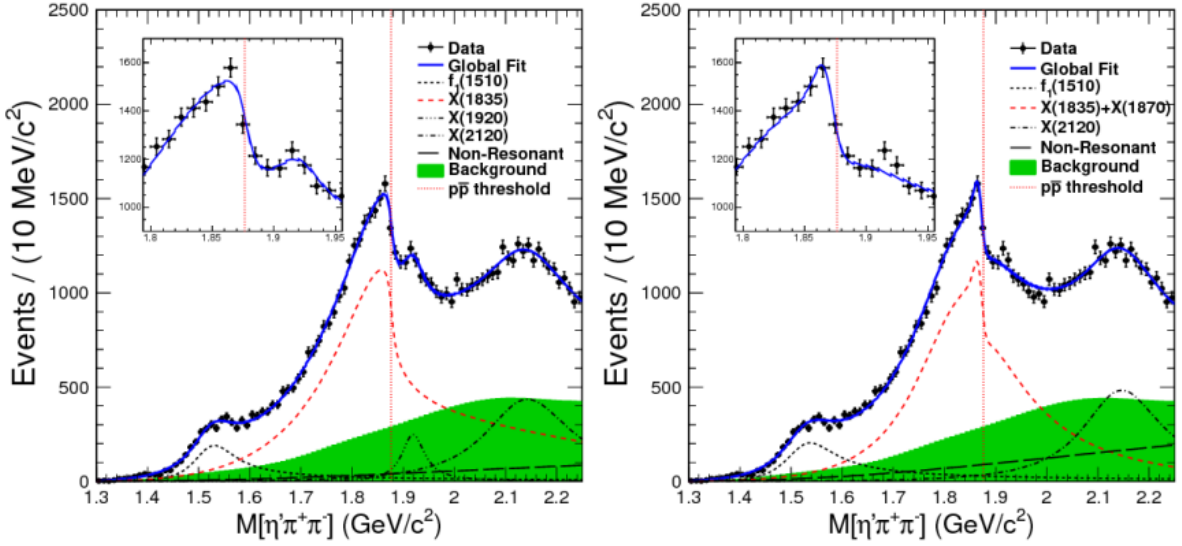}
\caption{Fit results based on the Flatt$\acute{\rm e}$ line-shape (left) and a coherent sum of two Breit-Wigner amplitudes. The position of $p\overline{p}$ mass-threshold is represented by a dashed vertical line, data by the points with error bars, signal and background contributions by various dashed curves, and the total fit by a solid blue curve.   }
\label{ppthreshold}    
\end{figure}

An anomalous enhancement $X(p\overline{p}$) at $p\overline{p}$ mass-threshold was originally observed by BESII in $J/\psi \to \gamma p \overline{p}$ decay~\cite{bes2ppbar}, and later confirmed by both BESIII~\cite{bes3ppbar} and CLEO experiments~\cite{cleoppbar}. This enhancement might be due to an $X(1835)$ state, which was first observed by BESII in $J/\psi \to \gamma \eta' \pi^+\pi^-$ decays~\cite{bes2x1835} and later confirmed by BESIII in the same process~\cite{bes3x1835} and $J/\psi \to \gamma \eta K_S^0K_S^0$ decays~\cite{bes3x1835KsKseta}. The observed spin-parity of $X(p\overline{p})$ and $X(3872)$ states is in favor of $J^P=0^-$~\cite{bes3ppbar,bes3x1835KsKseta}. By analyzing the decay process of $J/\psi \to \gamma \eta' \pi^+\pi^-$ with  1.1 billion $J/\psi$ events collected in 2012,  BESIII has observed a significant abrupt change in the slope of the $\eta' \pi^+\pi^-$ mass spectrum at the $p\overline{p}$ mass threshold~\cite{anamolousthreshatppbar}. Two typical models are used to characterize the $\eta' \pi^+\pi^-$ line-shape around 1.85 GeV/$c^2$ (Figure~\ref{ppthreshold}). The first  incorporates the opening of the decay threshold in the mass spectrum (Flatt$\acute{\rm e}$ formula), while the second  uses a coherent sum of two resonant amplitudes. Both the models describe  the data, and well, suggesting the existence of either a  $p\overline{p}$ molecule-like or bound state~\cite{anamolousthreshatppbar}.
\subsection{Observation of $\bm{\eta}$(1475) and X(1835) in $\bm{J/\psi \to \gamma \gamma \phi}$}
\begin{figure}[!htp]
 \sidecaption
\includegraphics[scale=.80]{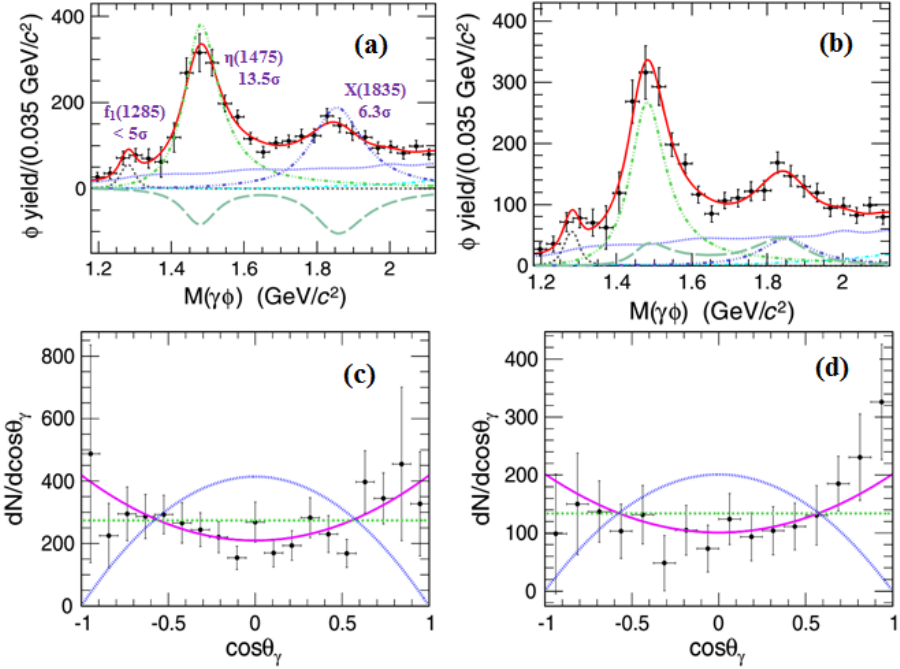}
\caption{(Top) fits to the $M_{\gamma \phi}$  distributions for the case of (a) constructive and (b) destructive interference, and (bottom) the efficiency corrected  $\cos\theta_{\gamma}$ distributions  for (c) $1.4 < M_{\gamma K^+K^-} < 1.6$ GeV/$c^2$ and (d) $1.75 < M_{\gamma K^+K^-} < 1.9$ GeV/$c^2$. The points with error bars are data, and solid and dashed curves represent the fit results.  In  $\cos\theta_{\gamma}$ distribution that is fitted by $1+ \alpha cos\theta_{\gamma}^2$, the solid pink, dashed blue and  dotted blue curves  correspond to the hypotheses  of angular distribution parameter $\alpha =1$, 0 and -1, respectively.}
\label{eta1475}     
\end{figure}

The $\eta(1440)$ is a puzzling state first observed in $p\overline{p}$ annihilation at rest into $\eta (1440) \pi^+\pi^-$, $\eta(1440) \to K\overline{K}\pi$~\cite{eta1440}, and later in the $J/\psi$ radiative decays to $K\overline{K}\pi$~\cite{eta1440kkpi}, $\gamma \rho$~\cite{eta1440grho} and $f_0(980)\pi^0$~\cite{eta1405}. Many experimental results reveal the existence of two different pseudoscalar states, the $\eta (1405)$ and  $\eta(1475)$~\cite{glueballeta1405}. The former is interpreted as an excellent candidate for $0^{-+}$ glueball~\cite{glueballeta1405}, and later as the first excitation of the $\eta'$. A triangle singularity is proposed to explain these anomalies by assuming both $\eta(1405)$ and $\eta(1475)$ to be a single state, the $\eta(1440)$, that appears as different line-shapes in different channels~\cite{trianglesing}. The $J/\psi \to \gamma \gamma \phi$ decay is studied using 1.3 billion $J/\psi$ events collected by the BESIII detector~\cite{eta1475X}. Two resonant structures corresponding to $\eta (1475)$ and $X(1835)$  are observed in the $\gamma \phi$ invariant mass spectrum with a significance of $13.5\sigma$ and $6.3\sigma$, respectively, by taking into account to interference between these two resonant structures during the fit (Figure~\ref{eta1475} (a,b)). The angular distributions of these resonances are observed to be $J^{PC} =0^{-+}$ (Figure~\ref{eta1475} (c,d)).  These results reveal that both $\eta(1475)$ and X(1835) contain a sizable $s\overline{s}$ component. 

\subsection{Observation of X(2370) in $\bm{J/\psi \to \gamma KK\eta'}$}
\begin{figure}[!htp]
\sidecaption
\includegraphics[scale=.62]{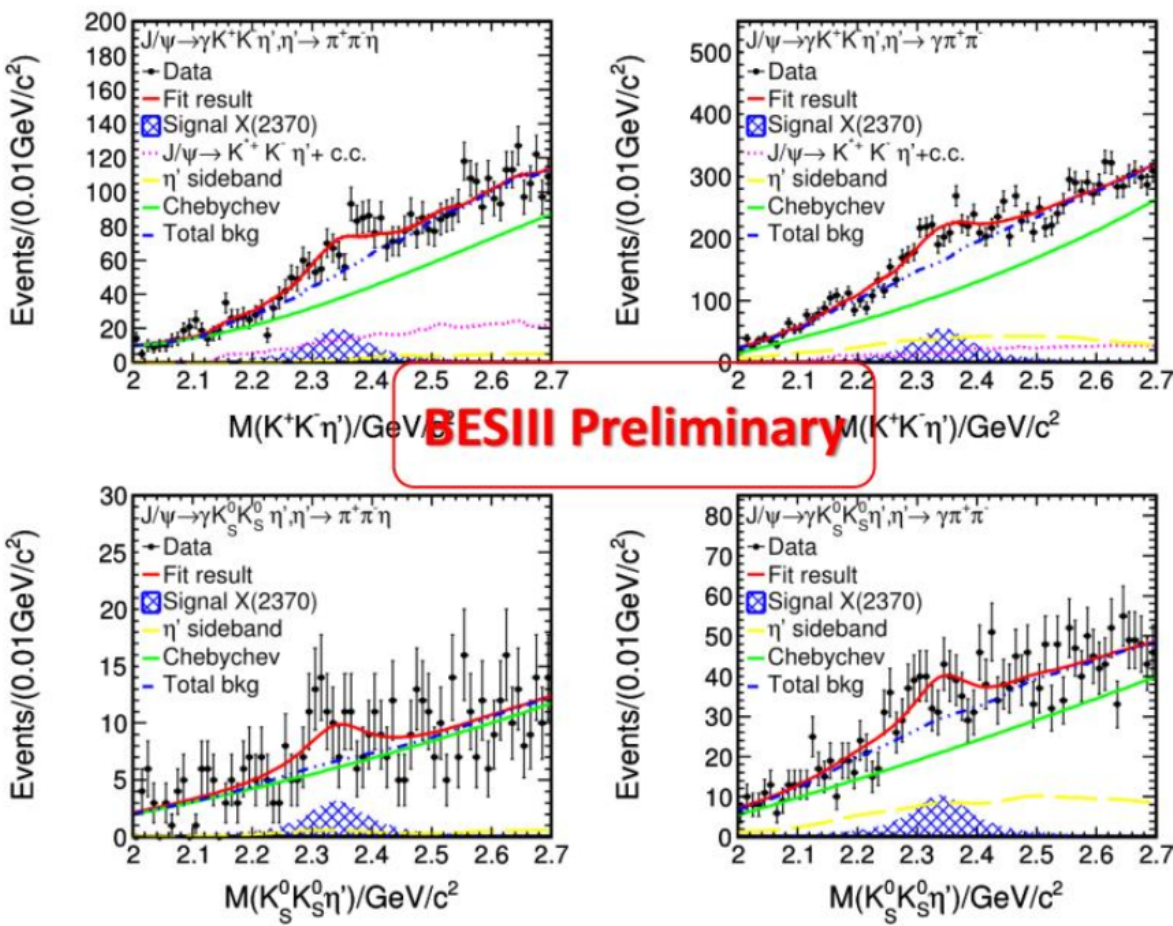}
\caption{Results of the fit for an X(2370) signal on the invariant mass spectra of $K^+K^-\eta'$ (top) and $K_S^0K_S^0 \eta'$ (bottom), where $\eta'$ decays to $\pi^+\pi^- \eta(\to \gamma \gamma)$ (left) $\gamma \rho(\to \pi^+\pi^-)$ (right). The points with error bars are data, the blue grid area represents the signal, the blue dashed double dotted curves show the total background contributions of $\eta'$ sideband (solid yellow curve), $J/\psi \to K^{*+}K^-\eta'$ (pink dashed curve) in $M(K^+K^-\eta')$ spectra only, and remaining backgrounds (solid green curve),  and solid  red  curves show the total fit result.    }
\label{kketp}    
\end{figure}

The $X(2120)$ and $X(2370)$ states are observed in the $\pi^+\pi^-\eta'$ invariant mass spectrum through the decay of $J/\psi \to \gamma \pi^+\pi^-\eta'$~\cite{bes2x1835,anamolousthreshatppbar}. The lattice QCD theory predicts the ratios of branching fractions of pseudoscalar glueball decays $\Gamma_{G\to KK\eta'}/\Gamma_G^{tot}$ and  $\Gamma_{G\to \pi\pi\eta'}/\Gamma_G^{tot}$ to be 0.011 and 0.090, respectively, where the glueball mass  is set at 2.37 GeV/$c^2$~\cite{Eshraim}. The observation of these states in $K \overline{K}\eta'$ decays would support the hypothesis that they are glueballs. The search for $X(2120)$ and $X(2370)$ is performed via the decays of $J/\psi \to \gamma K^+K^- \eta'$ and $J/\psi \to \gamma K_S^0K_S^0 \eta'$  using 1.3 billion $J/\psi$ events. A structure around 2.34 GeV/$c^2$, the X(2370), is observed in the $K\overline{K}\eta'$  spectra with a statistical significance of $7.6 \sigma$ (Figure~\ref{kketp}). The product branching fractions for $J/\psi \to \gamma X(2370),X(2370) \to K^+K^-\eta'$, and  $J/\psi \to \gamma X(2370),X(2370)\to K_S^0K_S^0 \eta'$ are determined to be $[1.86\pm0.39~(stat.)\pm0.29~(sys.)]\times10^{-5}$ and  $[1.19\pm0.37~(stat.)\pm0.18~(sys.)]\times10^{-5}$, respectively.  No evidence for the  X(2120) production is found, and $90\%$ confidence level (CL)  upper limits on product branching fractions for $J/\psi \to\gamma X(2120)\to K^+K^- \eta'$  and $J/\psi\to\gamma X(2120)\to K_S^0 K_S^0 \eta'$ are set at $1.48\times10^{-5}$ and $4.57\times10^{-6}$, respectively,  for the first time. These results are preliminary.

\section{Search for strangeonium-like states}
\subsection{Search for strangeonium-like structure $Z_S$  at 2.125 GeV}
\begin{figure}[!htp]
  \sidecaption
\includegraphics[scale=.20]{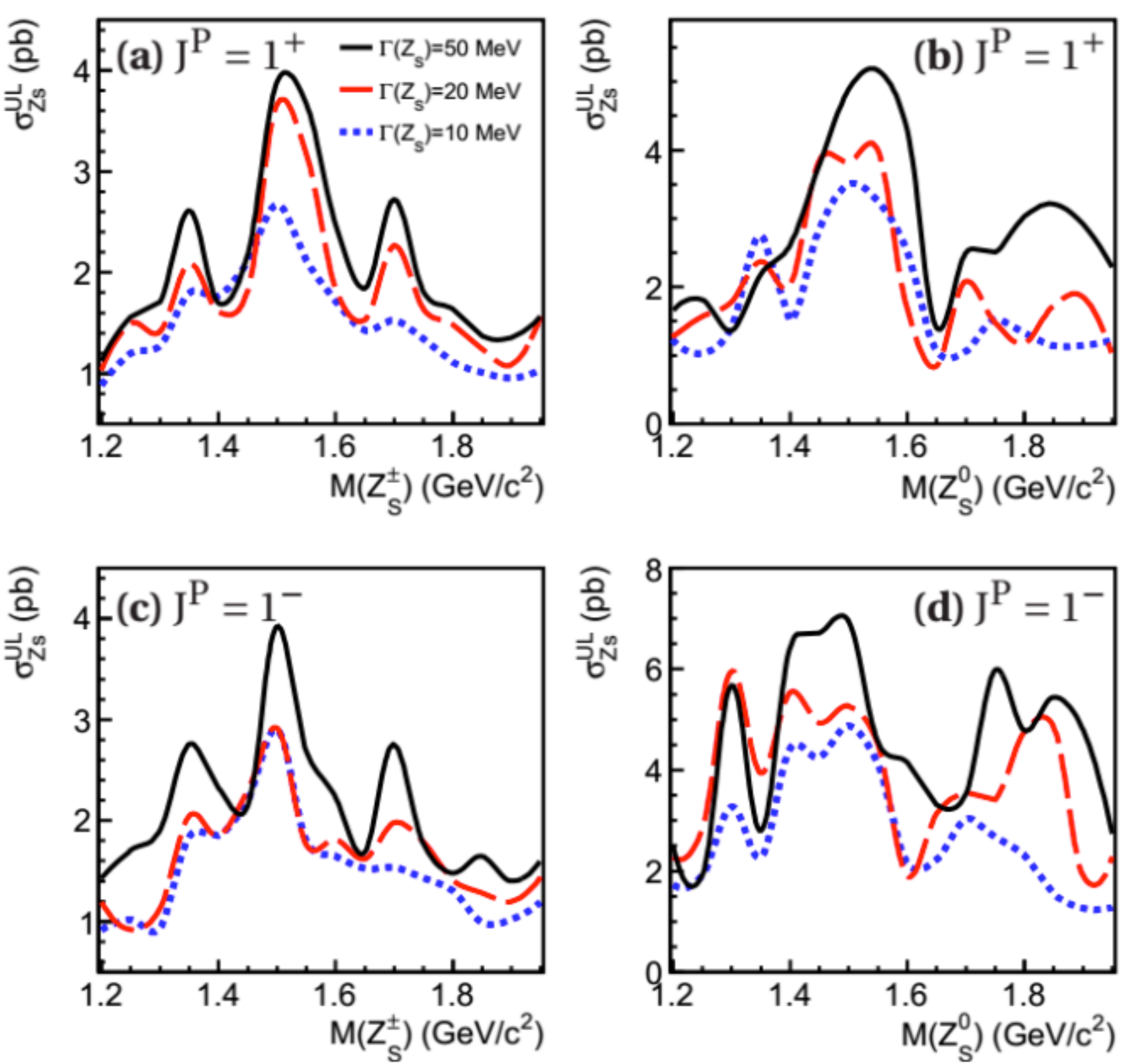}
\caption{ $90\%$  CL  upper limits on cross-section  for different widths of $Z_S$  as a function of  assumed $Z_S$ mass for the cases (a) $J^P=1^+$ of  $Z_S^{\pm}$,  (b) $J^P=1^+$ of $Z_S^0$, (c) $J^P=1^-$ of  $Z_S^{\pm}$ and  (d) $J^P=1^-$ of $Z_S^0$.  }
\label{ULzs}     
\end{figure}

A search for $Z_S$ strangeonium-like structure is performed in the process $e^+e^- \to \phi \pi\pi$ using a data sample corresponding to an integrated luminosity of $(108.49 \pm 0.75)$ pb$^{-1}$, taken at the CM energy of 2.125 GeV by the BESIII detector~\cite{zsbes3}.  A partial wave analysis  of $e^+e^- \to \phi \pi\pi$ is performed to describe the di-pion invariant spectrum after applying all the selection criteria. The fit includes the amplitudes of $e^+e^- \to \phi \sigma,~\phi f_0(980),~\phi f_0(1370)$ and $\phi f_2(1270)$ with a spin-parity of $J^P=1^+$. No evidence for $Z_S$ production is found in the invariant mass spectra of $\phi \pi^{\pm}$ and $\phi \pi^0$ around 1.4 GeV/$c^2$. The $90\%$ CL upper limits on the cross-section of $Z_S$ production are determined under different assumptions of mass, width and spin-parity of $Z_S$ (Figure~\ref{ULzs}). In addition, the cross-section of $e^+e^- \to \phi \pi^+\pi^-$ and $e^+e^- \to \phi \pi^0\pi^0$ at 2.125 GeV/$c^2$ are measured as ($343.0 \pm 5.1 \pm 25.1$) pb and ($208.3\pm 7.6 \pm 13.6$) pb, respectively. The first one slightly differs from BaBar~\cite{babarZs} and Belle~\cite{belley2175} measurements, being consistent within $3\sigma$, whereas the second is consistent with the BaBar measurement~\cite{babarZs} .

\subsection{Observation of $\bm{h_1}$(1380) in $\bm{J/\psi \to \eta' K \overline{\bm K}\pi}$}
\begin{figure}[!htp]
 \sidecaption
\includegraphics[scale=.80]{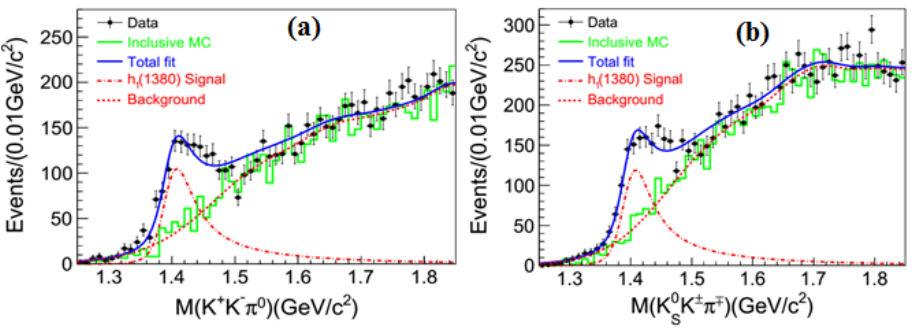}
\caption{Fits to  $K^+K^-\pi^0$ (left) and $K_S^0K^{\pm}\pi^{\mp}$ (right) mass spectra with interference between signal and background. Points with error bars are data, red dash-dotted and dashed curves are signal and background contributions, and solid blue curves show the total fit.}
\label{h1380}     
\end{figure}

The $h_1(1380)$ is considered to be a strangeonium state and an $s\overline{s}$ partner of $J^{PC} = 1^{+-}$ axial-vector state $h_1(1170)$. Experimentally, this state  was observed by the LASS~\cite{lass} and Crystal Barrel~\cite{cb} collaborations, and later confirmed by BESIII via $\psi(3686) \to \gamma \chi_{cJ}$, $\chi_{cJ} \to \phi h_1(1380)$, $h_1(1380) \to K^*(892) \overline{K}$ with $J=1,2$~\cite{bes3h1380}.  BESIII has recently reported the first observation of $J/\psi \to \eta' h_1(1380)$, where $h_1(1380) \to K^*(892)\overline{K} + c.c \to K^+K^-\pi^0/K_S^0K^{\pm}\pi^{\mp}$, using a sample of 1.3 billion $J/\psi$ events~\cite{h1380}. The $h_1(1380)$ resonance is observed in $K^+K^-\pi^0$ and $K_S^0K^{\pm}\pi^{\mp}$ mass spectra with a statistical significance larger than $10\sigma$ by performing a simultaneous fit (Figure~\ref{h1380}).  An isospin symmetry violation is found in $h_1(1380)$ decays between $h_1(1380) \to K^*(892)^+ K^- $ and $h_1(1380) \to K^*(892)^0 K^0 $. In addition, the mixing angle between the $h_1(1170)$ and $h_1(1380)$ is also determined to be $(35.9 \pm 2.6)^{\circ}$. This measured angle supports that the quark contents of the $h_1(1380)$ and $h_1(1170)$ are predominantly by $s\overline{s}$ and $u\overline{u}+d\overline{d}$, respectively.

\section{Summary and future prospects}
The BESIII has conducted a series of studies on gluonic and strangeonium-like states using the data samples collected at $J/\psi$, $\psi(3886)$ and $\phi(2170)$ resonances. Several  such particles are reported to have been observed recently. The BESIII will accumulate additional 9 billion $J/\psi$ events before the end of 2019, that will be utilized to improve the precision of measured states and possibly discover many new exotic states in the near future. 
\section{Acknowledgment}
This work has supported in part by the National Natural Science Foundation of China (NFSC) under contract No. 11705192.
\input{referenc}

\end{document}

%% file: referenc.tex
%
%
%

%% file: BESIIILHSforFPCP2018Proc.bbl
\begin{thebibliography}{99.}%
%
%
\bibitem{latticeQCD} Y. Chen {\it et al.},  Phys. Rev. D {\bf 73}, 014516 (2006).
\bibitem{reditJpsglueb}  L. C. Gui {\it et al.},  Phys. Rev. Lett. {\bf 110}, 021601 (2013); Y. B. Yang {\it et al.} Phys. Rev. Lett. {\bf 111}, 091601 (2013).
\bibitem{zc} M. Ablikim {\it et al.} (BESIII Collaboration), Phys. Rev. Lett. {\bf 110}, 252001 (2013);
 \bibitem{zcbelle} Z. Q. Liu {\it et al.} (Belle Collaboration), Phys. Rev. Lett. {\bf 110}, 252002 (2013).  
\bibitem{y4260} B. Aubert {\it et al.} (BaBar Collaboration), Phys. Rev. Lett. {\bf 95}, 142001 (2005); M. Ablikim {\it et al.} (BESIII Collaboration), Phys. Rev. Lett. {\bf 118}, 092001 (2017).
\bibitem{babary2175} B. Aubert {\it et al.} (BaBar Collaboration), Phys. Rev. D {\bf 74}, 091103(R) (2006).
\bibitem{belley2175}  C. P. Shen {\it et al.} (Belle Collaboration), Phys. Rev. D {\bf 80}, 031101(R) (2009).  
\bibitem{bes2y2175} M. Ablikim {\it et al.} (BES Collaboration), Phys. Rev. Lett. {\bf 100}, 102003 (2008);
\bibitem{bes3y2175} M. Ablikim {\it et al.} BESIII Collaboration, Phys. Rev. D {\bf 91}, 052017 (2015);  Observation of $e^+e^- \to \eta Y(2175)$ at center-of-mass energies above 3.7 GeV, arXiv:1709.04323 (2017).
\bibitem{bes2ppbar} J. Z. Bai {\it et al.} (BES Collaboration), Phys. Rev. Lett. {\bf 91}, 022001 (2003).  
\bibitem{bes3ppbar} M. Ablikim {\it et al.} (BESIII Collaboration), Chin. Phys. C {\bf 34}, 421 (2010); Phys. Rev. Lett. {\bf 108}, 112003 (2012).   
\bibitem{cleoppbar}  J. P. Alexander {\it et al.} (CLEO Collaboration), Phys. Rev. D {\bf 82}, 092002 (2010).
\bibitem{bes2x1835} M. Ablikim {\it et al.} (BES Collaboration), Phys. Rev. Lett. {\bf 95}, 262001 (2005).
\bibitem{bes3x1835} M. Ablikim {\it et al.} (BESIII Collaboration), Phys. Rev. Lett. {\bf 106}, 072002 (2011).
\bibitem{bes3x1835KsKseta} M. Ablikim {\it et al.} (BESIII Collaboration), Phys. Rev. Lett. {\bf 115}, 091803 (2015).
\bibitem{anamolousthreshatppbar}  M. Ablikim {\it et al.} (BESIII Collaboration), Phys. Rev. Lett. {\bf 117}, 042002 (2016).
\bibitem{eta1440} P. H. Baillon {\it et al.}, Nuovo Cimento A {\bf 50}, 393 (1967).
\bibitem{eta1440kkpi} D. L. Scharre et al., Phys. Lett. B {\bf 97}, 329 (1980).
\bibitem{eta1440grho} J. Z. Bai {\it et al.} (BES Collaboration), Phys. Lett. B {\bf 594}, 47 (2004).
\bibitem{eta1405}  M. Ablikim {\it et al.} (BESIII Collaboration), Phys. Rev. Lett. {\bf 108}, 182001 (2012).  
\bibitem{glueballeta1405} L. Faddev, A. J. Niemi and U. Wiedner, Phys. Rev. D {\bf 70}, 114033 (2004).
\bibitem{trianglesing} J. J. Wu, X. H. Liu, Q. Zhao and B. S. Zou, Phys. Rev. Lett. {\bf 108}, 081803 (2012). 
\bibitem{eta1475X} M. Ablikim {\it et al.} (BESIII Collaboration), Phys. Rev. D {\bf 97}, 051101(R) (2018).
\bibitem{Eshraim}  W. I. Eshraim, S. Janowski, F. Giacosa and D. H. Rischke, Phys. Rev. D {\bf 87}, 054036 (2013).
\bibitem{zsbes3} M. Ablikim {\it et al.} (BESIII Collaboration), Phys. Rev. D {\bf 99}, 011101(R) (2018).
\bibitem{babarZs} J. P. Lees {\it et al.} (BaBar Collaboration), Phys. Rev. D {\bf 86}, 012008 (2012).
\bibitem{lass} D. Aston {\it et al.} (LASS Collaboration), Phys. Lett. B {\bf 201}, 573 (1988).  
\bibitem{cb} A. Abele {\it  et al.} (Crystal Barrel Collaboration), Phys. Lett. B {\bf 415}, 280 (1997).
\bibitem{bes3h1380} M. Ablikim {\it et al.} (BESIII Collaboration), Phys. Rev. D {\bf 91}, 112008 (2015).
\bibitem{h1380} M. Ablikim {\it et al.} (BESIII Collaboration),  Phys. Rev. D {\bf 98}, 072005 (2018).

\end{thebibliography}
